\documentclass[ journal]{IEEEtran}
\usepackage{url,amsmath,graphicx}
\usepackage{fancyhdr}
\usepackage{cite}
\usepackage{graphicx}
\usepackage{psfrag}
\usepackage{subfigure}
\usepackage{url}
\usepackage{stfloats}
\usepackage{amsmath}
\usepackage{array}
\usepackage{fancyhdr}
\usepackage{epsfig}
\usepackage{amssymb}
\usepackage{color}

\begin{document}
\title{  Unified Analysis and Optimization of D2D Communications in Cellular Networks Over Fading Channels  }
\author{Im\`ene~Trigui, Member, \textit{IEEE},  and~Sofi\`ene~Affes, Senior Member, \textit{IEEE},
 }

 \maketitle

  \begin{abstract}
This paper develops an innovative approach to the modeling and analysis of downlink cellular networks with  device-to-device (D$2$D) transmissions. The analytical embodiment of the signal-to-noise and-interference
ratio (SINR) analysis in general fading channels is unified due to the H-transform theory, a taxonomy  never considered before in stochastic geometry-based cellular network modeling and analysis.
 The proposed
framework has the potential, due to versatility of the Fox's H functions, of significantly simplifying the cumbersome analysis
procedure and representation of  D$2$D and cellular coverage, while subsuming  those previously derived for all the known simple and composite
fading models. By harnessing its tractability, the developed statistical machinery is employed
to launch an investigation into the  optimal design of coexisting  D$2$D and cellular communications.
We propose  novel coverage-aware power control combined with opportunistic access control to maximize the area spectral efficiency (ASE) of D$2$D communications.
Simulation results substantiate  performance gains achieved by the proposed optimization framework  in terms
of cellular communication coverage probability, average D$2$D
transmit power, and the ASE of D$2$D communications under different fading models and  link- and
network-level dynamics.
\let\thefootnote\relax\footnotetext{Work supported by the Discovery Grants and the CREATE PERSWADE programs of NSERC, and a Discovery Accelerator Supplement (DAS) Award from NSERC.}
 \end{abstract}
\section{Introduction}
The recent  sky-rocketing  data
demand has compelled both industry and
regulatory bodies to come up with new paradigm-shifting technologies able to keep pace with such stringent requirements and cope with the massive connectivity characterizing  future $5$G networks. Currently
being touted as a strong contender for $5$G networks \cite{5G}, \cite{5G2}, device-to-device (D$2$D) communications  allow direct
communication between cellular mobiles, thus bypassing
the network infrastructure, resulting
in shorter transmission distances and improved data rates than
traditional cellular networks.

In the past few years, D$2$D-enabled networks have been actively studied by the research
community.  For example, in \cite{d2}, it was shown that by allowing radio signals to be relayed by mobiles,  D$2$D communications can improve spectral efficiency and
the coverage of conventional cellular networks. Additionally, D$2$D has  been applied to machine-to-machine (M$2$M) communications \cite{machine} and proposed as a possible enabler of vehicle-to-vehicle (V$2$V) applications \cite{d4}. More recent
works \cite{and}-\!\!\cite{sawy} have modeled the user locations with PPP
 distributions and analytically tackled
D$2$D communication by harnessing  the powerful  stochastic geometry tools.

Notwithstanding these advances, computing the SINR in randomly deployed networks, namely D$2$D-enabled cellular networks, has been successfully tractable only for fading channels
and transmission schemes whose equivalent per-link power
gains follow a Gamma distribution with integer shape parameter (\!\!\cite{trigui} and references therein), while much work has  been achieved on evaluating the
performance of D$2$D networks over Rayleigh fading channels \cite{and}.
Such particular fading distributions have very often limited  legitimacy according to \cite{kmu},\!\cite{beaulieu}, who argued that these fading
models may fail to capture  new and more realistic fading environments. This is particularly true as new
communication technologies accommodating a wide range of usage scenarios with diverse link requirements are continuously being introduced
and analyzed, for example,  body-centric and  millimeter-wave communications.
Recently few works have been conducted to consider D$2$D networks with general fading channels \cite{rices}-\!\cite{d2dsku1}. However, besides being channel-model-dependent, these works  relied  on  series representation
methods (e.g., infinite series in \cite{sd} and Laguerre polynomial series in \cite{d2dsku},\!\cite{d2dsku1} ) thereby expressing the interference functionals as an infinite series of
higher order derivative terms given by the Laplace transform of the interference power. These methods cannot lend themselves to closed-form expressions and, hence, require complex numerical evaluation.

 For the successful coexistence of
D$2$D and cellular users, efficient interference management, e.g., through power or access
control, is required. Recently, extensive research on power allocation strategies aiming to maximize the spectral efficiency of D$2$D communication
in random network models were studied and analyzed \cite{p1}-\!\cite{p3}. In \cite{p1}, channel-aware power control algorithms aiming to maximize the D$2$D sum rate
are proposed and  analyzed using stochastic geometry. In \cite{p2},  SIR-aware access scheme  based on the conditional coverage  probability of
D$2$D underlaid cellular networks  is proposed to increase the aggregate rate of D$2$D links. Similarly, \cite{p3} proposes to enhance the sum rate of D$2$D links
by optimally finding groups and access probabilities. To the
best of the authors' knowledge, none of these works consider generalized fading channels when proposing  access and power control  schemes
to accommodate multiple D$2$D pairs underlaid in a cellular network.  Yet, these works only focused on the simplistic Rayleigh fading.

In this paper, we focus on the design of access
control and power allocation strategies
for D$2$D communication underlaying  wireless networks under generalized fading conditions,  an uncharted
territory wherein the throughput potential of such
networks remains unquantified. The contributions of this paper are as follows:

\begin{itemize}

\item  We propose an analytical framework
based on newly established tools from
stochastic geometry analysis \cite{trigui} to evaluate the  cellular and D$2$D SINR distributions in general fading conditions embodying the H-transform theory. We establish extremely useful results for the SINR and interference distributions never reported previously in the literature.

\item We successfully unify our analysis  framework in the sense that it
can be applied for any fading channel whose
envelope follows the form of $x^{\beta} e^{-\lambda x}{\rm H}(c x^{\zeta})$, where $H(\cdot)$ stands for the Fox's H function \cite{H1}, e.g.,
Nakagami-$m$,  Weibull, or $\kappa$-$\mu$ and shadowed $\kappa$-$\mu$  to account for various small-scale
fading effects such as LOS/NLOS (line-of-sight/non-LOS) conditions,
multipath clustering,  composite fading in specular
or inhomogeneous radio propagation, and power imbalance between
the in-phase and quadrature signal components.

\item In order to guarantee the coverage probability of cellular users in a distributed manner, we derive the  interference budget of a typical D$2$D link, which represents
the allowable transmit power level of D$2$D transmitters. We formulate an optimization
problem to find the access probability which maximizes
 the average area spectral efficiency  utility of D$2$D communication
underlaying multiple cells subject to cellular
coverage-aware power budget. The proposed opportunistic
access requires only statistical CSI (channel state information), in contrast to the
centralized resource allocation which requires full CSI, thereby  inducing less delay in the network.

\item We derive  simple expressions
for the optimal access probability and D$2$D
coverage-aware power budget based on an approximation of the D$2$D coverage probability  under both Nakagami-$m$ and Weibull fading channels. The developed machinery is prone to handle  more comprehensive fading models, namely the shadowed $\kappa$-$\mu$.

\end{itemize}

The remainder of this paper is organized as follows. We
describe the system model in Section II. In Section III, we put forward the fading model-free statistical distribution
of cellular and D$2$D links in Section III, then we put forth
 the unifying H-transform analysis over the considered fading channels.  We exploit  in Section VI the developed statistical machinery   to present  the cellular coverage-aware power control and ASE of
D$2$D communications under different fading conditions.  We present numerical results in Section V and
conclude the paper in Section VII with some closing remarks.

\section{System Model}

We envision a D2D-enabled cellular network model
in which
the locations of macro BSs (MBS) and D$2$D users are
distributed according to the independent homogeneous PPPs (HPPPs) $\Psi_c$ and $\Psi_d$
with intensities $\lambda_c$ and $\lambda_d$, respectively.  We assume (i) all users are served by the MBS  from which they receive the
strongest  average power as their serving stations,  which is equivalent
to the nearest BS association criterion, and (ii) that a typical user is allowed to connect to a randomly selected  D$2$D transmitter (Tx).
It is worth mentioning that the methodology of analysis
to be presented later in this paper can be applied to different
techniques pertaining to relaxing assumption (ii). It is worth mentioning here that the new analysis methodology proposed in this paper can be applied to different
techniques pertaining to relaxing assumption (ii)  by considering  content availability \cite{dil}, proximity \cite{saha1} and clustering \cite{saha},\cite{dil2}.
Due to lack of space, we  delegate for the sake of clarity these stand-alone extension materials to  future works.
We assume that all links between the transmitters (BS and
D$2$D Tx) to the typical user undergo distance dependent pathloss
and small-scale fading. Then, the received power  from the MBS/D$2$D Tx located at $Z_x \in \Psi_x$, $x = \{c,d\}$ is given as
$P(Z_x )= P_{x} h_{x} \|Z_x \|^{-\alpha},$  where $P_c$ and $P_d$ are the transmit powers of the MBS and D$2$D Txs, respectively, $\alpha\geq 2$ is the pathloss exponent and  $\{h_x\}$ is an i.i.d. sequence of
 random variables  modeling the channel power.
 The
signal-to-interference plus noise ratio (SINR)  at the location of the typical
user  when connecting to the nearest MBS ($\mathrm{SINR}_x$) or to a randomly chosen  D$2$D Tx ($\mathrm{SINR}_d$) can be expressed as
\begin{equation}
\mathrm{SINR}_x=\frac{P_x h_x \|Z_x^{*} \|^{-\alpha} }{I_x+\sigma^{2}}, \quad x = \{c,d\}
\end{equation}
where, $\|Z_x^{*} \|$ is the distance to the nearest MBS ($Z_c^{*}=\arg\max_{Z_c \in\Psi_{c}} P_c \|Z_c \|^{-\alpha}$)
 and to the random D$2$D Tx, respectively.
The  co-channel interference from other D$2$D Txs  or from  other MBS  are denoted by $I_x=\sum_{ Z_x \in\Psi_{x} \backslash {Z_x^{*}}}P_{x} h_{x}\|Z_x \|^{-\alpha}$.

\section{Generalized SINR Analysis}
In this section we derive the SINR distribution of a typical user in (i) a stand-alone cellular network in which only MBSs  are available to provide
 coverage to any user, (ii) a stand-alone D$2$D network in which only D$2$D devices are available to
provide  coverage to any user. The obtained statistical machinery will be harnessed later to investigate underlaid D$2$D networks.

\textit{Theorem 1:}
The  SINR complementary
cumulative distribution function (CCDF) of
D$2$D and cellular links, defined as $\mathbb{P}^{x}(T)\triangleq \mathbb{P}\left(\mathrm{SINR}_x=\frac{P_x h_x r^{-\alpha}}{I_x+\sigma^{2}}\geq T\right)$  for $x\in\{ d , c\}$, is given by
\begin{eqnarray}
\mathbb{P}^{x}(T)&=&\frac{1}{T}\int_{0}^{\infty}\underbrace{{\rm \cal{E}}_h\!\!\left[h~ {\rm H}_{1,2}^{1,0}\!\left[\!\frac{  h \xi}{T}\left|
\begin{array}{ccc}(0,1) \\ (0,1),(-1,1) \end{array}\right.\right]\!\right]}_{\Psi(\xi,T)}\nonumber \\ && \times {\cal E}_r\left[\exp\left(-\frac{\sigma^{2}}{P_x}\xi r^{\alpha} -  {\cal A}^{x}\!\left(\xi r^{\alpha}, \delta\right)\right)\right]d\xi,
\label{GF}
\end{eqnarray}
where $r=\|Z_x^{*} \|$, $x\in\{ d , c\}$, $\delta=\frac{2}{\alpha}$,  ${\cal E}_z[.]$ is the expectation with respect to the random
variable $z$,  ${\rm H}_{a,b}^{c,d}[\cdot]$  denotes the  Fox-H function \cite{matai}, \cite{H1}   and
\begin{equation}
\left\{
  \begin{array}{ll}
  {\cal A}^{d}(\xi, \delta)= \pi  \lambda_d \xi^{\delta}\Gamma\left(1-\delta\right) {\cal E}\left[h^{\delta}\right] , & \hbox{\!\!\!\! D$2$D;} \\
    {\cal A}^{c}(\xi, \delta)=\frac{\pi \delta \lambda_c \xi ~{\cal E}_{h}\left[ h~{\rm {}_{2}F_{2}}\left(\! 1\!-\!\delta,1;2\!-\!\delta,2; - \frac{\xi h}{ r^{2/\delta}}\! \right)\right] }{ r^{2(1/\delta-1)}(1-\delta)}, & \hbox{\!\!\!\! Cellular.}
  \end{array}
\right.
\label{Ax}
\end{equation}
whereby  ${\rm {}_{p}F_{q}}(\cdot)$ and $\Gamma(\cdot)$ stand for the generalized hypergeometric  \cite[Eq. (9.14.1)]{grad} and the incomplete gamma  \cite[Eq.(8.310.1)]{grad} functions, respectively.

\textit{Proof:}  See Appendix A for details.\\
Theorem 1 demonstrates the general expressions of the Laplace transforms of $I_d$ and $I_c$
 as well as the D$2$D and cellular SINR CCDFs without assuming any specific random channel gain and distance models\footnote{In this paper,
the unbounded pathloss model is used due to its  mathematical tractability. However more realistic models notably bounded
pathloss models (BPM) ($(1+d)^{-\alpha}$, $\min(1, r^{-\alpha}$) can be studied through the general SINR expression in Theorem 1.}.

Notice that $\Psi(\xi,T)$ in (\ref{GF}) is an  integral transform that
involves the Fox's H-function as kernel, whence called H-transform.  The
H-transforms involving Fox's H-functions as kernels were
first suggested by Verma \cite{verma} with the help of ${\cal L}_2$-theory for
integral transforms in the Lebesgue space ${\cal L}_2$ \cite{verma}.
So far, integral transforms, such as the classical  Laplace,
Mellin, and Hankel transforms have been used
successfully in solving many problems pertaining to stochastic geometry modeling in cellular networks (cf.\cite{and} and \cite{trigui} and references therein).
 However, to the best of our knowledge, this paper is the first to introduce the Fox's-H function and  H-transforms to cellular network analysis.
Since H-functions subsume most of the known special functions
including Meijer's G-functions \cite{matai},  then by virtue of the essential so-called Mellin
 operation, involved in the Mellin transform of two H-functions,   $\Psi(\xi,T)$ culminate in a H-function
for any channel model with probability density function (PDF) $f_h(\cdot)$.

 A  single H-variate PDF considers homogeneous
radio propagation conditions and captures composite effects of multipath
fading and shadowing, subsuming
most of typical models such as Rayleigh, Nakagami-$m$,
Weibull, $N$-Nakagami-$m$, (generalized) ${\cal K}$-fading,
and Weibull/gamma fading \cite{simon} as its special
cases.

In contrast, to characterize specular
and/or inhomogeneous environments, the multipath component
consists of a strong LOS or specularly-reflected wave as
well as unequal-power or correlated in-phase and quadrature
scattered waves \cite{kmu}, \cite{simon}, \cite{paris}. Another class of
H-variate (degree-$2$) PDF that is the product of an exponential function and a Fox's H-function is used to account for specular or
inhomogeneous radio propagation conditions including a variety of relevant models
such as Rician, $\kappa$-$\mu$, Rician/LOS gamma, and  $\kappa$-$\mu$/LOS gamma (or $\kappa$-$\mu$ shadowed)  fading \cite{kmu},\cite{paris} as
special cases.

In this paper,  we choose, however, to work with
single H-variate fading PDFs to keep the presentation as compact
as possible. Some other fading models that can still be considered within the framework of this paper are
degree-$2$ H-variate fading models including the $\kappa$-$\mu$ and the shadowed $\kappa$-$\mu$. We illustrate this fact
in Appendix D.

\textit{Proposition 1 (Nakagami-$m$ Fading):} The D2D and cellular SINR CCDFs over
 Nakagami-$m$ fading are given by
\begin{eqnarray}
\mathbb{P}_{m}^{x}(T)\!\!\!\!&=&\!\!\!\!\frac{\pi \delta \lambda_x \left(\frac{\sigma^{2}}{P_x}\right)^{\delta/2}}{\Gamma( m)}\!\!\!\int_{0}^{\infty}\frac{{\rm H}_{1,1}^{0,1}\left[\frac{ \Omega \xi}{ m T}\left|
\begin{array}{ccc}(1-m,1)\\(0,1) \end{array}\right. \right]}{\xi^{1+\frac{\delta}{2}}}\nonumber \\ && \!\!\!\!\!\!\!\!{\rm H}_{1,1}^{1,1}\!\left[\pi \lambda_x \left(1+{\cal Q}_x\right) \frac{\left(\frac{\sigma^{2}}{P_x}\right)^{\delta}}{\xi^{\delta}} \left|
\begin{array}{ccc}(1-\delta,\delta) \\(0,1) \end{array}\right. \right]d\xi,
\label{COVmf}
\end{eqnarray}
where $x\in\{c,d\}$, and
 \begin{eqnarray}
\left\{
  \begin{array}{ll}
   {\cal Q}_d=\xi^{\delta}\left(\frac{\Omega}{ m}\right)^{\delta}\frac{\Gamma(1-\delta)\Gamma(m+\delta)}{\Gamma(m)},  \\
     {\cal Q}_c=\frac{ \delta \xi  \Omega~ {\rm {}_{3}F_{2}}\left( 1-\delta,1, m+1;2-\delta,2; - \frac{\Omega\xi }{m } \right)}{ (1-\delta)}.
  \end{array}
\right.
\label{Amf}
\end{eqnarray}

\textit{Proof:}  See Appendix B for details.

\textit{Definition 1:} Consider the Fox's-H function ${\rm H}_{p,q}^{m,n}\left[ x \left|\!\!\!
\begin{array}{ccc} (a_1,A_1), \ldots, (a_n,A_N) \\ (b_1,b_1), \ldots, (b_n,b_N) \end{array}\right. \!\!\!\right]$ defined by \cite[Eq. (1.1.1)]{kilbas}. It's
asymptotic expansion near $x = \infty$ is given by \cite[Eq. (1.5.9)]{kilbas} as
\begin{equation}
{\rm H}_{p,q}^{m,n}(x)\underset{x\rightarrow\infty}{\approx} \eta x^{d},
\label{Has1}
\end{equation}
where $d= \max \left(\frac{a_i-1}{A_i}\right), i=1,\ldots,n$ and $\eta$ is calulated as in  \cite[Eq. (1.5.10)]{kilbas}.

\textit{Corollary 1 (Limits of Network Densification in Nakagami-m Fading) :} The downlink SINR
 saturates past a certain network density as
\begin{equation}
\!\!\!\!\!\!\lim_{\lambda_c\rightarrow\infty} \mathbb{P}_{m}^{c}(T)=\frac{ \left(\frac{P_c }{ \sigma^{2}}\right)^{\delta/2}}{\Gamma(m)}\!\!\int_{0}^{\infty}\!\frac{{\rm H}_{1,1}^{0,1}\left[\frac{ \Omega \xi}{ m T}\left|
\begin{array}{ccc}(1-m,1) \\(0,1) \end{array}\right. \right]}{\xi^{1-\frac{\delta}{2}}\left(1+{\cal Q}_c\right)}dx.
\label{COVmI}
\end{equation}

\textit{Proof:}
Applying (\ref{Has1}) to (\ref{COVm}) when $\lambda_c\rightarrow\infty$ yields the result after recognizing that $d=-1$ and $\eta=\frac{1}{\delta}$.

Corollary 1  proves that  at some point ultra-densification will no longer be able to deliver
significant coverage gains. Although, some other works have also identified such fundamental scaling regime
for network densification \cite{and},\!\!\cite{trigui}, its  network performance limits in terms of coverage  have never been exactly quantified as in Corollary 1.

Due to potentially high density of devices, D$2$D networks are overwhelmingly interference-limited. In this respect, the D$2$D SIR distribution becomes
  \begin{equation}
\mathbb{P}_{m}^{d,\sigma\simeq0}(T)\overset{(a)}{=}\frac{1}{\Gamma(m)} {\cal E}_{r_d}\left[ {\rm H}_{1,2}^{2,0}\!\left[ \lambda_d \pi \kappa_m T^{\delta} r^{2}\left|\!\!\!
\begin{array}{ccc} (1,\delta) \\ (0,1), (m,\delta) \end{array}\right. \!\!\!\right]\right],
\label{COVdm}
\end{equation}
where $(a)$  follows form (\ref{COVm}) when $\sigma^{2}\simeq0$  by resorting to \cite[Eq.(2.19)]{matai} with $\kappa_m=\frac{\Gamma(1-\delta)\Gamma(m+\delta)}{\Gamma(m)}$.  The analytical result in (\ref{COVdm}) applies
to any spatial distribution of $r_d$. It derives under Rayleigh distance, using \cite[Eq.(2.19)]{matai}, as
\begin{equation}
\mathbb{P}_{m}^{d, \sigma\simeq0}(T) =\frac{1}{\Gamma(m)} {\rm H}_{2,2}^{2,1}\!\left[\kappa_m T^{\delta} \left|
\begin{array}{ccc} (0,1),(1,\delta) \\ (0,1), (m,\delta) \end{array}\right.\right].
\end{equation}


\textit{Proposition 2 (Weibull Fading):} The Weibull fading channel accounts for the nonlinearity of a
propagation medium
with a physical fading parameters $\nu$.   When $h$ follows  a Weibull distribution with parameters $(\nu, \Phi=\Omega^{\nu})$ \cite{simon}, then the  SINR CCDFs
of D$2$D and cellular links are
\begin{eqnarray}
\mathbb{P}_{\cal W}^{x}(T)\!\!\!&\!\!\!\!\!=\!\!\!\!\!&\!\!\!\frac{\pi \delta \lambda_x \left(\frac{\sigma^{2}}{P_x}\right)^{\delta/2}\nu T^{\nu}}{  \Phi} \int_{0}^{\infty}\frac{{\rm H}_{1,1}^{1,0}\!\left[ \left(\frac{ T}{ \xi }\right)^{\nu}\frac{1}{\Phi}\left|\!\!\!
\begin{array}{ccc} (1-\nu,\nu) \\ (0,1)\end{array}\right. \!\!\!\right]}{\xi^{\nu+1+\frac{\delta}{2}}}\nonumber \\ && \times {\rm H}_{1,1}^{1,1}\!\left[\pi \lambda_x \left(1+{\cal G}_x\right) \frac{\left(\frac{\sigma^{2}}{P_x}\right)^{\delta}}{\xi^{\delta}} \left|
\begin{array}{ccc}(1-\delta,\delta) \\(0,1) \end{array}\right. \right]d\xi,
\label{COVw1}
\end{eqnarray}
where $x\in\{c,d\}$, and
\begin{equation}
\left\{
  \begin{array}{ll}
   {\cal G}_{d}= \xi^{\delta}\Phi^{\frac{\delta}{\nu}}\Gamma\left(1-\delta\right)\Gamma\left(1+\frac{\delta}{\nu}\right), \\
     {\cal G}_{c}=\frac{\delta}{1-\delta} {\rm H}_{3,3}^{3,1}\left[\frac{1}{\xi \Phi^{\frac{1}{\nu}} }\left|
\begin{array}{ccc}\!\!(-1,1), (1-\delta,1),(1,1)\\ (1,\frac{1}{\nu}),(-\delta,1), (0,1)\end{array}\right. \!\!\!\right].
  \end{array}
\right.
\label{Axwf}
\end{equation}

\textit{Proof:} See Appendix C for details.

\textit{Corollary 2 (Limits of Network Densification in Weibull Fading):} For any SINR target $T$, the cellular coverage probability in Weibull fading
 flattens out starting from some network density $\lambda_c$  as
\begin{eqnarray}
\lim_{\lambda_c\rightarrow\infty} \mathbb{P}_{{\cal W}}^{c}(T)&=&\frac{ \nu T^{\nu}\left(\frac{P_c }{ \sigma^{2}}\right)^{\delta/2}}{\Phi} \nonumber \\ && \!\!\!\!\!\!\!\!\!\!\!\!\!\!\!\!\!\!\!\!\!\!\!\!\!\!\!\!\!\!\!\!\!\!\!\!\!\!\!\!\!\!\!\!\!\!\!\!\!\!\!\!\!\int_{0}^{\infty}\!\frac{\xi^{\frac{\delta}{2}-\nu-1}{\rm H}_{1,1}^{1,0}\!\left[ \left(\frac{ T}{ \xi }\right)^{\nu}\frac{1}{\Phi}\left|\!\!\!
\begin{array}{ccc} (1-\nu,\nu) \\ (0,1)\end{array}\right. \!\!\!\right]}{1+\frac{\delta}{1-\delta} {\rm H}_{3,3}^{3,1}\left[\frac{1}{\xi \Phi^{\frac{1}{\nu}} }\left|
\begin{array}{ccc}\!\!(-1,1), (1-\delta,1),(1,1)\\ (1,\frac{1}{\nu}),(-\delta,1), (0,1)\end{array}\right. \!\!\!\right]}d\xi.
\label{COVwI}
\end{eqnarray}

\textit{Proof:} The result follows in the same line of (\ref{COVmI}) while using (\ref{COVw1}).

The D$2$D SIR distribution is obtained from (\ref{COVw}) as
 \begin{eqnarray}
\mathbb{P}_{\cal W}^{d, \sigma\simeq0}(T)\!\!\!&\overset{(a)}{=}& \!\!\!{\cal E}_r\left[{\rm H}_{1,2}^{2,0}\!\left[\pi \lambda_d \kappa_{\cal W} T^{\delta} r^{2}\left|\!\!\!
\begin{array}{ccc} (1,\delta) \\ (0,1), (1,\frac{\delta}{\nu}) \end{array}\right. \!\!\!\right]\right],
\label{COVwi}
\end{eqnarray}
where $(a)$ follows form applying \cite[Eqs. 2.3]{H1} whereby $\kappa_{\cal W}=\Gamma(1-\delta)\Gamma(1+\frac{\delta}{\nu})$.

\textit{Remark 1:} The Rayleigh fading is a special case of  (\ref{COVm})  and (\ref{COVw}) when $m=1$ and $\nu=1$, respectively, thereby yielding
\begin{eqnarray}
\mathbb{P}^{x}(T)&\overset{(a)}{=}& \pi \delta \lambda_x \left(\frac{\sigma^{2}}{P_x}\right)^{\delta/2}\nonumber \\ && \!\!\!\!\!\!\!\!\!\!\!\!\!{\rm H}_{1,1}^{1,1}\!\left[\pi \lambda_x \left(1+{\cal R}_x\right) \frac{\left(\frac{\sigma^{2}\Omega}{P_x}\right)^{\delta}}{T^{\delta}} \left|\begin{array}{ccc}(1-\delta,\delta) \\(0,1) \end{array}\right. \right],
\label{pcovr2}
\end{eqnarray}
with
 \begin{equation}
\left\{
  \begin{array}{ll}
   {\cal R}_d= \frac{\pi T^{\delta}}{\delta \sin(\pi \delta)},  \\
    {\cal R}_d=\frac{T \delta}{1-\delta}  {\rm {}_{2}F_{1}}\left( 1-\delta,1;2-\delta; - T \right),
  \end{array}
\right.
\label{Ar}
\end{equation}
where $(a)$ follows after recognizing that ${\rm H}_{1,1}^{1,0}\left[x\left|
\begin{array}{ccc} 1\\1\end{array}\right. \right]=\delta\left[x\!-\!1\right]$, when $\delta[x]$ stands for the DiracDelta function, i.e., $\delta[x]=0/; x\neq0$.
The coverage formulas in (\ref{pcovr2}) matches the well-known major results for Rayleigh fading obtained in \cite[ Theorem 1]{and} with the valuable add-on  of being in closed-form.

\section{Area Spectral Efficiency Optimization}

Hereafter, we consider a cellular network
underlaid with D$2$D Txs. We assume an ALOHA-type channel access for both D$2$D and cellular Txs with probability $p_d$  and $p_c$, respectively.
 Then, the set of active MBS/D$2$D Txs also forms a HPPP $\Psi_i^{\{TX\}}$ with density $\lambda_i p_i$ where $i\in\{c,d\}$.
We assume that each cellular transmitter has
its intended receiver at a fixed distance $r_c$ in a random direction.
Similarly, each D$2$D receiver is located at distance
$r_d$ from its corresponding transmitter.

The ASE, often referred to as
network throughput, is a measure of the number of users
that can be simultaneously supported by a limited radio frequency
bandwidth per unit area.

\textit{Definition 2:}
 In D$2$D-underlaid cellular networks, the ASE of D$2$D communications
can be expressed as \cite{p2}
\begin{equation}
{\cal T}(T_d)=p_d \lambda_d {\mathbb P}(\mathrm{SINR}_d(P_d)>T_d)\log_2(1+T_d),
\label{aread}
\end{equation}
where  ${\mathbb P}(\mathrm{SINR}_d>T_d)={\mathbb P}^{d}(T_d)$ is the mean of the D$2$D coverage probability (previously derived in Section III),
$p_d \lambda_d$ denotes the effective D$2$D link density without any inactive D$2$D link, and
$p_d$ is the  access probability.


In what follows, capitalizing on the statistical framework developed in section III, we present quality-based power control strategies for D$2$D communications under both Nakagami-$m$ and Weibull fading channels aiming to reduce the interference caused by D$2$D Txs and to maximize the ASE of D$2$D communications.

\subsection{Cellular Coverage Probability-Aware
Power Control}
 Consider an arbitrary cellular transmitter $x \in \Psi_c^{\{TX\}}$ and its
associated receiver at distance $r_c$.  We are interested in investigating the joint effect of the
interference coming from the surrounding BS and the D$2$D
Txs on the downlink coverage probability while ignoring the noise
by assuming $\sigma^{2}=0$. Therefore the SIR at the cellular receiver is
expressed as
\begin{eqnarray}
\mathrm{SIR}_c(P_d)\!\!\!\!&=&\!\!\!\!\frac{ h_{c}r_c^{-\frac{2}{\delta}}}{\sum_{i\in  \Psi_c^{\{TX\}}\backslash\{0\}}h_ir_{i}^{-\frac{2}{\delta}}\!+\!\eta \sum_{j\in \Psi_d^{\{TX\}}} h_{j}r_{j}^{-\frac{2}{\delta}}}\nonumber \\ && \!\!\!\!\!\!\!\!=\frac{ h_{c}r_c^{-\frac{2}{\delta}}}{I_c+\eta I_d},
\label{sirc}
\end{eqnarray}
where $\eta=\frac{P_d}{P_c}$ is the ratio of the transmit powers of the D$2$D.
Let $\rho^{c}_{th}$ be the operator-specified cellular success probability
threshold. The maximum transmit power for D$2$D Txs  is obtained
by solving the following optimization problem:
\begin{eqnarray}
\max \quad \quad P_d \nonumber \\  && \!\!\!\!\!\!\!\!\!\!\!\!\!\!\!\!\!\!\!\!\!\!\!\!\!\!\!\!\!\!\!\!\!\!\!\!\!\!\!
\text{s.t.} \quad \quad {\mathbb P}\left(\mathrm{SIR}_c(P_d)>T_c\right)\geq \rho^{c}_{th},
\label{qosc}
\end{eqnarray}
where for  any $P_d$,  the
cellular coverage probability, conditioned on $r_c$, can be
expressed over general fading  based on (\ref{sirc}) and employing (\ref{GF}) as
\begin{eqnarray}
{\mathbb P}\left(\mathrm{SIR}_c(P_d)>T_c\right)=\mathbb{P}^{c}(T_c)|_{r_c}\!\!&=&\!\!\nonumber \\ &&\!\!\!\!\!\!\!\!\!\!\!\!\!\!\!\!\!\!\!\!\!\!\!\!\!\!\!\!\!\!\!\!\!\!\!\!\!\!\!\!\!\!\!\!\!\!\!\!\!\!\!\!\!\!\!\!\!\!\!\!\!\!\!\!\!\!\!\!\!\!\!\!\!\!\!\!\!\!\!\!\!\!\!\!\!\!\!\!\!\!\frac{1}{T_c}\int_{0}^{\infty}\!\!\!\!\Phi(\xi, T_c)\exp\!\left(-  {\cal A}^{c}\!\left(\xi r_c^{2}, \delta\right)\!-\!\eta^{\delta} {\cal A}^{d}\!\left(\xi r_c^{2}, \delta\right)\!\right)\!d\xi,
\label{ccov}
\end{eqnarray}
where $\Phi(\xi, T_c)$ is a function of only  the fading parameters as previously shown in (\ref{COVm}) and (\ref{COVw}). Moreover, assuming that the distance-based association policy imposes no constraint on the location of interfering MBSs to the probe receiver, we have
\begin{equation}
{\cal A}^{x}\!\left(\xi , \delta\right)=\pi p_x \lambda_x {\cal E}[h_x^{\delta}] \Gamma\left(1-\delta\right) \xi^{\delta}, \quad x\in\{c,d\},
\end{equation}

\subsubsection{D$2$D Power Control Under Nakagami-$m$ Fading}
In this section, we assume that all  links experience Nakagami-$m$
flat fading channel. The fading severity of the Nakagami-$m$
channel is captured by the parameter $m_d$ for all links originating
from the D$2$D transmitters, while the fading severity of
the cellular communication and interference links is captured
by  the parameter $m_c$.

\textit{Proposition 3:}
The maximum D$2$D transmit power can be expressed under Nakagami-$m$ fading as
\begin{eqnarray}
\!\eta p_d^{\frac{1}{\delta}}&=&\max\Bigg\{\Bigg(\frac{-p_c \lambda_c \kappa^{c}_{m}}{ p_d \lambda_d \kappa^{d}_m \left(\!\frac{m_c}{P_c m_d}\!\right)^{\delta}}+\nonumber \\ &&\frac{(1-m_c){\cal W}\bigg(\!\frac{\left(\rho^{c}_{th}\Gamma(m_c)\right)^{\frac{1}{m_c-1}}}{m_c-1}\bigg)}{\pi T^{\delta}_c  r_c ^{2}p_d\lambda_d \kappa^{d}_m \left(\!\frac{m_c}{P_c m_d}\!\right)^{\delta}}\Bigg)^{\frac{1}{\delta}}, 0\Bigg\},
\label{MAP1}
\end{eqnarray}
which acts as an average individual interference budget of
each D$2$D Tx to guarantee the coverage probability of cellular users, where $\kappa^{x}_{m}=\frac{ \Gamma(1-\delta)\Gamma(m_x+\delta)}{\Gamma(m_x)}$, $x\in \{d,c\}$ and ${\cal W}(\cdot)$ is the principal branch of the Lambert function \cite{grad}.

\textit{Proof:}
The cellular coverage probability under Nakagami-$m$ fading follows from (\ref{ccov}) as
\begin{eqnarray}
\mathbb{P}^{c}_{m}(T_c)|_{r_c}&\!\overset{(a)}{=}\!&\frac{1}{\Gamma(m_c)} \nonumber \\ && \!\!\!\!\!\!\!\!\!\!\!\!\!\!\!\!\!\!\!\!\!\!\!\!\!\!\!\!\!\!\!\!\!\!\!\!\!\!\times {\rm H}_{1,2}^{2,0}\!\left[ T_c^{\delta}r_c^{2}\!\left(p_d\lambda_d \kappa^{d}_m \left(\frac{m_c}{m_d}\right)^{\delta} \eta^{\delta}\!\!+ \!\!p_c \lambda_c \kappa^{c}_{m}\right)\!\!\left|\!\!\!
\begin{array}{ccc} (1,\delta) \\ (0,1), (m_c,\delta) \end{array}\right. \!\!\!\!\right]\nonumber \\
&\!\!\!\!\!\!\!\!\!\!\!\!\!\!\!\!\!\!\!\!\!\!\!\!\!\!\!\!\!\!\!\!\!\!\!\!\!\!\!\!\!\!\!\!\!\!\!\!\!\!\overset{(b)}{\underset{T_c\rightarrow\infty}{\approx}}&\!\!\!\!\!\!\!\!\!\!\!\!\!\!\!\!\!\!\!\!\!\!\!\!\!\!\!\!\!\frac{T_c^{\delta(m_c-1)}r_c^{2}\left(p_d\lambda_d \kappa^{d}_m \left(\frac{m_c}{m_d}\right)^{\delta} \eta^{\delta}\!\!+ \!\!p_c \lambda_c \kappa^{c}_{m}\right)^{m_c\!-\!1}}{\Gamma(m_c)}\!\nonumber \\ &&\!\!\!\!\!\!\!\!\! e^{-T_c^{\delta}r_c^{2}\left(p_d\lambda_d \kappa^{d}_m \left(\frac{m_c}{m_d}\right)^{\delta} \eta^{\delta}+ p_c \lambda_c \kappa^{c}_{m}\right)},
\label{covmm}
\end{eqnarray}
where $(a)$ follows from (\ref{COVdm}) while ${\cal A}_m^{d}$ is given in (\ref{Am}) and
${\cal A}_{m}^{c}={\cal A}_{m}^{d}\!\left(m_d \leftarrow m_c, p_d\leftarrow p_c \right)$.

\textit{Definition 3:} Consider the Fox's-H function defined by \cite[Eq. (1.1.1)]{kilbas}. Its
asymptotic expansion near $x = \infty$ when  $n=0$ is given by \cite[Eq. (1.7.14)]{kilbas}
\begin{equation}
{\rm H}_{p,q}^{q,0}(x)\sim x^{\frac{\nu+\frac{1}{2}}{\Delta}} \exp\left[-\Delta \left(\frac{x}{\rho}\right)^{1/\Delta}\right],
\label{Has}
\end{equation}
where $\nu$, $\Delta$, and $\rho$ are constants defined in \cite[Eq. (1.1.8)]{kilbas}, \cite[Eq. (1.1.9)]{kilbas}, and \cite[Eq. (1.1.10)]{kilbas}, respectively. Then $(b)$ follows after recognizing that $\delta=1$, $\rho=1$, $\nu=m_c-\frac{3}{2}$. Thus, the problem in (\ref{qosc}) can be solved by (\ref{covmm})(b)-$\rho^{c}_{th}=0$. The latter is a homogeneous equation, which is solvable, thereby yielding the desired result after some mathematical manipulations.

\subsubsection{D$2$D Power Control Under Weibull Fading}

In this section, we assume that all  links experience Weibull
flat fading channel. Specifically, the cellular interfering link suffers from the Weibull ($\nu_c$) fading and the D$2$D interference link
experiences the Weibull ($\nu_d$) fading.

\textit{Proposition 4:} The  cellular success probability-aware
power control  under Weibull fading  yields
\begin{equation}
\!\!\!P_d^{*}=\max\left\{\left(-\frac{P_c^{\delta} p_c \lambda_c \kappa^{c}_{\cal W}}{p_d \lambda_d \kappa^{d}_{\cal W}}+\frac{ P_c^{\delta} \rho_c\left(-\frac{\ln(\rho^{c}_{th})}{\sigma_c}\right)^{1/\sigma_c}}{\pi p_d \lambda_d \kappa^{d}_{\cal W} T_c^{\delta}r_c^{2}}\right)^{\frac{1}{\delta}}\!\!, 0\right\},
\label{MAP2}
\end{equation}
where $\kappa^{x}_{\cal W}=\Gamma(1-\delta)\Gamma(1+\frac{\delta}{\nu_x})$, $x\in\{d,c\}$, $\sigma_c=1+\delta(1-\frac{1}{\nu_c})$, and $\rho_c=\delta^{-\delta}\left(\frac{\delta}{\nu_c}\right)^{\delta/\nu_c}$.

\textit{Proof:}
The cellular coverage probability  under Weibull fading follows from (\ref{ccov})  as
\begin{eqnarray}
\mathbb{P}^{c}_{{\cal W}}(T_c)|_{r_c}&\overset{(a)}{=}&\nonumber\\ && \!\!\!\!\!\!\!\!\!\!\!\!\!\!\!\!\!\!\!\!\!\!\!\!\!\!\!\!\!\!\!\!\!\!\!\!\!\!\!\!\!\!\!\!\!\!\!\!{\rm H}_{1,2}^{2,0}\left[T_c^{\delta}r_c^{2}\left(p_d \lambda_d \kappa^{d}_{\cal W} \eta^{\delta}+ p_c \lambda_c \kappa^{c}_{\cal W}\right)\left|
\begin{array}{ccc} (1,\delta) \\ (0,1), (1,\frac{\delta}{\nu_c}) \end{array}\right. \right],\nonumber \\
&\underset{T_c\rightarrow\infty}{\overset{(b)}{\approx}}& \!\!\!\!\!\!\!\!\!e^{-\sigma_c\left(\frac{T_c^{\delta}r_c^{2}\left(p_d \lambda_d \kappa^{d}_{\cal W} \eta^{\delta}+ p_c \lambda_c \kappa^{c}_{\cal W}\right)}{\rho_c}\right)^{1/\sigma_c}},
\label{COVwd}
\end{eqnarray}
where  $(a)$ follows from applying (\ref{COVw}) whereby ${\cal A}_{\cal W}^{d}$ is given in (\ref{Axw}) and
${\cal A}_{\cal W}^{c}={\cal A}_{\cal W}^{d}\!\left(\nu_d \leftarrow \nu_c, p_d\leftarrow p_c \right)$. Moreover $(b)$ follows from resorting to (\ref{Has}).
Subsequently, by solving (\ref{COVwd})(b)-$\rho^{c}_{th}=0$, we get (\ref{MAP2}) after some mathematical manipulations.

In reality, the individual interference budgets in (\ref{MAP1}) and (\ref{MAP2}) may
be calculated and broadcast by MBSs to each D$2$D Tx at the
initial stage. In order to use the licensed spectrum, each D$2$D
Tx must obey the individual interference budget in its power
allocation stage. Note that our problem formulation is based on the
distributed power control framework where cellular users and D$2$D Txs do not need to
share location or  channel state, which implies that the individual interference budget in (\ref{MAP1}) and (\ref{MAP2})
do not require the instantaneous CSI which is in fact  difficult
to get accurately especially upon high mobility
of cellular and D$2$D users. Under the proposed distributed
power allocation framework,  a D$2$D Tx selects its transmit power
based solely on the knowledge of the cross-tier communication distance $r_c$, the users and MBS spatial density, and the joint effect of path loss and fading.
Compared to most existing schemes for D$2$D power control that are based on
the real-time CSI to mitigate interference (\!\!\!\cite{p1} and references therein), the proposed power control framework, being statistically featured, does not burden the network latency.

In particular, from (\ref{MAP1}) and (\ref{MAP2})
we prove that the cellular user coverage probability guarantee
can be distributively satisfied regardless of whether the D$2$D transmitters adapt their transmit power or access probability.
 In other words, D$2$D users can
tune either of these two parameters representing an  interference budget
that each D$2$D pair may not exceed toward the cellular users.

\subsection{D$2$D ASE-Aware Access Probability}

In densely deployed D$2$D networks, both D$2$D to cellular and inter-D$2$D interferences  would be very high. As a result, the cellular coverage
probability threshold may not be guaranteed, even under individual D$2$D interference
budget, especially when the target cellular SIR threshold is high. Hereafter,  instead of allowing all D$2$D transmitters to access the
channels, a part of D$2$D pairs cannot access the network to decrease
interferences. Hence we propose  to  extend the cellular coverage probability-aware
power control by integrating it with opportunistic access control
to maximize the area spectral efficiency of D$2$D communications
while decreasing both inter-D$2$D and cross-tier interferences.

After obeying to the individual interference budget in its power allocation stage,  each D$2$D Tx  maximizes its
ASE utility ${\cal T}(T_d)$ by optimizing  the access probability $p_d$. We formulate the individual access-aware
design problem of D$2$D Tx as
\begin{eqnarray}
\underset{p_d}{\max}\quad {\cal T}(T_d)\nonumber \\
\text{s.t.} \quad 0<p_d\leq1,
\label{as1}
\end{eqnarray}
where ${\cal T}(T_d)$ is defined in (\ref{aread}) with maximum permissible transmit power for
an arbitrary  D$2$D user at a particular MAP $p_d$  obtained from (\ref{MAP1}), and $\mathbb{P}^{d}(P_d, p_d)$ is the SIR coverage probability of an arbitrary D$2$D link  obtained as
\begin{eqnarray}
\!\!\!\!\!\!\mathbb{P}^{d}(T_d)|_{r_d}&\triangleq&\mathbb{P}\left(\mathrm{SIR}_d=\frac{h_{d}r_d^{-\frac{2}{\delta}}}{\eta^{-1}I_c+I_d}\geq T_d\right)\nonumber \\ &&\!\!\!\!\!\!\!\!\!\!\!\! \!\!\!\!\!\!\!\!\!\!\!\!\!\!\!\!\!\!\!\!\!\!\!\!\!\!=\frac{1}{T_d}\!\!\int_{0}^{\infty}\!\!\!\!\!\! \!\Phi(\xi, T_d)\exp\!\left(-  \eta^{-\delta}{\cal A}^{c}\!\left(\xi r_d^{2}, \delta\right)\!-\! {\cal A}^{d}\!\left(\xi r_d^{2}, \delta\right)\!\right)\!d\xi.
\label{dcov}
\end{eqnarray}

\subsubsection{D$2$D ASE under Nakagami-$m$ Fading}
Under Nakagami-$m$ fading, and conditioned on $r_d$, the SIR coverage probability of an arbitrary D$2$D link under transmit power adaptation (i.e., considering (\ref{MAP1})) follows from applying (\ref{dcov}) while considering  (\ref{COVdm}) as
\begin{eqnarray}
\mathbb{P}^{d}_{m}(T_d)|_{P_d^{*}}&=&\frac{1}{\Gamma(m_d)}\nonumber \\ && \!\!\!\!\!\!\!\!\!\!\!\!\!\!\times {\rm H}_{1,2}^{2,0}\!\left[ p_d\lambda_d \kappa^{d}_m T_d^{\delta}r_d^{2}\Xi\left|\!\!\!
\begin{array}{ccc} (1,\delta) \\ (0,1), (m_d,\delta) \end{array}\right. \!\!\!\!\right]\nonumber \\&\!\!\!\!\!\!
\!\!\!\!\!\!\!\!\!\!\!\!\!\!\!\!\!\!\!\!\!\!\!\!\!
\!\!\!\!\!\!\!\!\!\!\!\!\!\!\!\!\!\!\!\overset{(a)}{\underset{T_d\rightarrow\infty}{\approx}}&\!\!\!\!\!\!
\!\!\!\!\!\!\!\!\!\!\!\!\!\!\!\!\!\!\!\frac{\left(p_d \lambda_d \kappa^{d}_m T_d^{\delta}r_d^{2}\Xi\right)^{m_d-1}}{\Gamma(m_d)}\!e^{-p_d\underbrace{\lambda_d \kappa^{d}_m T_d^{\delta}r_d^{2}\Xi}_{\Xi_1}},
\label{pdm}
\end{eqnarray}
where $P_d^{*}$  is the maximum D$2$D transmit power of D$2$D Tx identified by cellular success probability-aware power control,
\begin{equation}
\Xi=\left(1-\frac{p_c\lambda_c \kappa^{c}_{m} T_c^{\delta}r_c^{2}}{(1-m_c){\cal W}\left(-\frac{(\rho^{c}_{th}\Gamma(m_c))^{\frac{1}{m_c-1}}}{m_c-1}\right)}\right)^{-1},
\end{equation}
and $(a)$ follows from  applying the algebraic
asymptotic expansions of the Fox's-H function in (\ref{Has}) with several mathematical manipulations.\\
Based on (\ref{pdm}), (\ref{as1}) can be transformed further to
\begin{eqnarray}
\underset{p_d}{\max}\quad\quad p_d^{m_d}e^{-p_d \Xi_1}\nonumber \\
\text{s.t.} \!\!\!\!\!\!\!\quad\quad 0<p_d\leq1,
\label{as2}
\end{eqnarray}
where $\Xi_1=\kappa^{d}_m T_d^{\delta}r_d^{2}\Xi$.

\textit{Proposition 5:} The optimal access probability $p_d^{*}$
under Nakagami-$m$ fading verifies
\begin{equation}
\frac{\partial p_d^{m_d}e^{-p_d \Xi_1}}{\partial p_d}=0,
\end{equation}
and is easily decided,  after some manipulations, by
\begin{equation}
p_d^{*}=\min\left\{\frac{m_d}{\lambda_d \kappa^{d}_m T_{d}^{\delta}r_d^{2}\Xi}, 1\right\}.
\label{pdm}
\end{equation}
 The  D$2$D area
spectral efficiency when operating at ($P_d^{*}$, $p_d^{*}$)
can be quantified under Nakagami-$m$ fading as
\begin{equation}
{\cal T}^{*}_{d}|_{p_d^{*}}\approx\frac{m_d^{m_d}e^{-m_d}\log_2(1+T_d)}{\Gamma(m_d) \kappa^{d}_m T_d^{\delta}r_d^{2}\Xi},
\label{tpdm}
\end{equation}
where $e \simeq0.277$.

\subsubsection{D$2$D ASE under Weibull Fading}
 The area spectral efficiency of D$2$D underlay cellular networks under Weibull fading and cellular success probability-aware
power control (i.e., considering (\ref{MAP2})) can be expressed as
\begin{equation}
{\cal T}_d|_{p_d}\approx p_d \lambda_d \log_2(1+T_d)\exp\left(\!\!\!-\sigma_d\!\!\left(p_d \underbrace{\frac{\lambda_d \kappa^{d}_{\cal W}T_d^{\delta}r_d^{2}\Pi}{\rho_d}}_{\Pi_1}\right)^{\!\!1/\sigma_d}\right),
\label{areaweibull}
\end{equation}
where $\sigma_d=1+\delta(1-\frac{1}{\nu_d})$,  $\rho_d=\delta^{-\delta}\left(\frac{\delta}{\nu_d}\right)^{\delta/\nu_d}$, and
\begin{equation}
\Pi=\left(1-\frac{\lambda_c \kappa^{c}_{{\cal W}} T_c^{\delta}r^{2}_c}{\rho_c\left(-\frac{\ln(\rho^{c}_{th})}{\sigma_c}\right)^{1/\sigma_c}}\right)^{-1}.
\label{pi}
\end{equation}
\textit{Proof:} Following the same rationale to obtain (\ref{pdm}), while considering (\ref{MAP2}),  yield the success probability of D$2$D underlay network under Weibull fading with maximum permissible transmit power. Plugging the obtained result into (\ref{aread}) completes the proof.

\textit{Proposition 6:} The optimal access probability ($p^{*}_d$)
which maximizes the
area spectral efficiency for D$2$D
underlay network under cellular success probability-aware power control  operating over Weibull fading  verifies
\begin{equation}
1-(\Pi_1 p_d)^{1/\sigma_d}=0, \quad 0<p_d\leq1,
\end{equation}
obtained from the maximization of the area spectral
efficiency in (\ref{areaweibull}), thereby yielding
\begin{equation}
p_d^{*}=\min\left\{\frac{\rho_d \left(1-\frac{\lambda_c \kappa^{c}_{{\cal W}} T_c^{\delta}r_c^{2}}{\rho_c\left(-\frac{\ln(\rho^{c}_{th})}{\sigma_c}\right)^{1/\sigma_c}}\right)}{\lambda_d \kappa^{d}_{\cal W} T_d^{\delta}r_d^{2}},1\right\}.
\label{pdw}
\end{equation}


Plugging $p_d^{*}$ into (\ref{areaweibull}) yields  the D$2$D underlay network ASE with cellular success probability-aware
power control and  opportunistic access control under Weibull fading as
\begin{equation}
{\cal T}^{*}_d|_{p_d^{*}}\approx\frac{\rho_d e^{-\sigma_d}\log_2(1+T_d)}{\kappa^{d}_{\cal W} T_d^{\delta}\Pi}=\frac{\rho_d e^{-\sigma_d}\log_2(1+T_d)}{\Gamma(1-\delta)\Gamma(1+\frac{\delta}{\nu_d})T_d^{\delta}r_d^{2}\Pi}.
\label{tpdw}
\end{equation}

Note that both the optimal access probability and ASE are  inversely proportional to the D$2$D link distance $r_d$.  That is implying that a D$2$D Tx with a short communication distance
has a higher access probability than a D$2$D transmitter with
a larger communication distance, because it has a potentially
higher SIR. We also notice that  $p_d^{*}$
is inversely related to the density
of D$2$D users ($\lambda_d$). Notice  that in many studies that intrinsically rely on
the optimality of access or power against density adaptation \cite{mark}, a similar behavior was noticed. However, in this
paper, we show that both the access and transmit power adaptations
by themselves are sub-optimal. To overpass such suboptimality, we extend cellular success probability-aware
power control by integrating it with opportunistic access control
to maximize the area spectral efficiency of D$2$D communications.

Note that (\ref{tpdm}) and  (\ref{tpdw}) coincide when $m_d=m_c=1$ and $\nu_c=\nu_d=1$ corresponding to the Raleigh case.

%

\section{Numerical And  Simulation Results}
In this section, numerical examples are shown to substantiate
the accuracy of the new unified mathematical framework and to explore from our new analysis the effects of both the link- and network- level dynamics\footnote{Link-level dynamics correspond to the uncertainty experienced due to
multi-path propagation and topological randomness, while  network-level
dynamics are shaped by medium access control, device/BS density, etc.} on the ASE of the underlay D$2$D
 network.

%
%
%

\begin{figure}[h!]
\centering
\centerline{\includegraphics[ width=8cm]{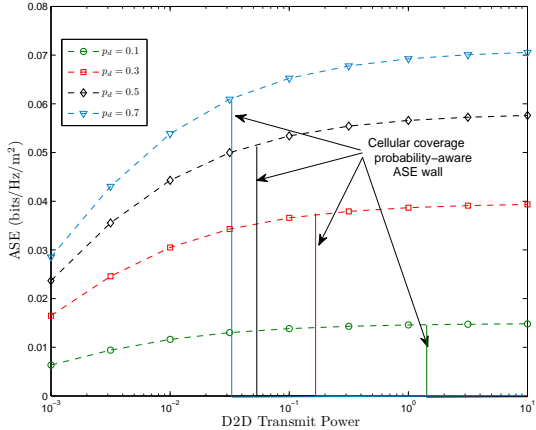}}
\caption{ASE of a D$2$D underlay network with transmit
power control under Nakagami-$m$ fading with $\lambda_d = 10^{-2}$, $\lambda_c = 10^{-3}$, $P_c = 1$, $\alpha=4$, $r_c = r_d = 1$,
$\rho^{c}_{th}= 0.1$, $p_c = 0.4$, $m_c = m_d = 1.5$, $T_c=5$ dB, and $T_d=3$ dB.}
\end{figure}

\begin{figure}[h!]
\centering
\centerline{\includegraphics[ width=8cm]{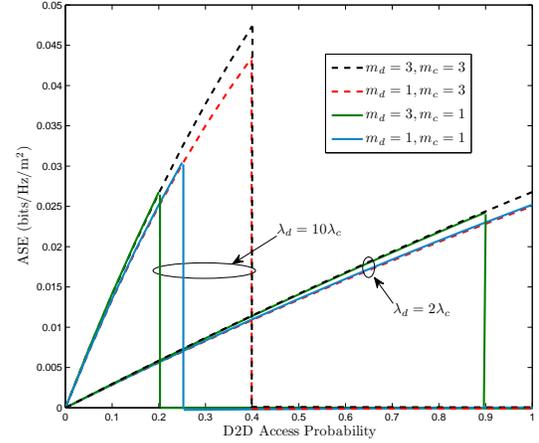}}
\caption{ ASE of a D$2$D underlay network with access control under Nakagami-$m$ fading  with $\lambda_c = 10^{-3}$, $P_c = 1$, $P_d=0.1$, $\alpha=4$, $r_c=r_d=1$,
$\rho^{c}_{th}= 0.1$, $p_c = 0.4$,  $T_c=5$ dB, and $T_d=3$ dB.}\label{fig:BPSK}	
\end{figure}

\begin{figure}[h!]
\centering
\centerline{\includegraphics[ width=8cm]{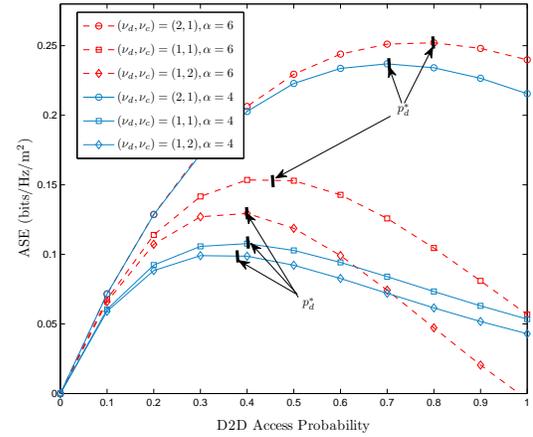}}
\caption{ Impact of  the path-loss
exponent on the ASE of the D$2$D underlay network under Weibull fading with
$\lambda_d = 10^{-2}$, $\lambda_c = 10^{-3}$, $P_c = 1$,  $r_c = r_d = 1$,
$\rho^{c}_{th}= 0.1$, $p_c = 0.4$, $T_c=5$ dB,  and $T_d=3$ dB.}\label{fig:BPSK}	
\end{figure}
\begin{figure}[h!]
\centering
\centerline{\includegraphics[ width=8cm]{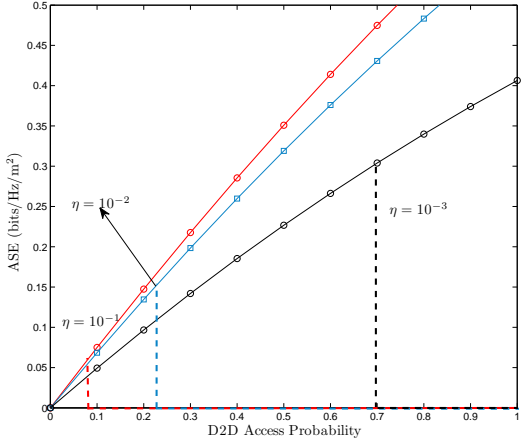}}
\caption{ ASE  under Weibull fading for various
value of $\eta$ with $\lambda_d = 10^{-2}$, $\lambda_c = 10^{-3}$, $P_c = 1$, $\alpha=4$, $r_c = r_d = 1$,
$\rho^{c}_{th}= 0.1$, $p_c = 0.4$, $\nu_c = \nu_d = 2.5$, $T_c=5$ dB,  and $T_d=3$ dB.}\label{fig:BPSK}	
\end{figure}

Fig.~1 shows the ASE of the D$2$D underlay network under Nakagami-$m$ fading
as a function of the D$2$D transmit power.  We
can notice that  D$2$D transmitters
can increase their transmit power to improve D$2$D
receivers' ASE up to a maximum value beyond which the operation becomes unfeasible due to the bound enforced by the cellular network.
Fig.~1 also shows that transmit power and access probability play a dual role. Indeed, an increase in the operational access probability inflicts
a higher co-channel interference to the cellular user and, hence,  a more
stringent operational constraint by a reduction in  the individual D$2$D interference budget (the maximum permissible
transmit power) and thereby of the ASE. Hence the gain obtained due
to an increase in the simultaneous transmissions may vanish because of the
reduction in the success probabilities of the individual links. This indicates that
there may exist an optimal operational point where the reduction in the link
coverage can be balanced by increasing the number of concurrent transmissions.

Fig.~2  plots the ASE of the D$2$D underlay network under Nakagami-$m$ fading
as a function of the D$2$D mean access probability. We observe that the maximum permissible density of the active
D$2$D transmitters is bounded due to the cellular user's
coverage constraint, thereby consolidating the trends of Fig.~1. Fig.~2 further investigates the  impact of  cellular and D$2$D channel fading severities and user densities
 on the ASE. For almost equally densely deployed cellular and D$2$D networks, D$2$D performance is governed by the fading severity $m_c$  rather than $m_d$. In this case, cellular users employ higher transmit power thereby
 bounding the D$2$D underlay network
performance due to  the inflicted cellular  interference. The dominant fading severity parameter is reversed when $\lambda_d> \lambda_c$
which is  hardly surprising because the increased density limits the D$2$D
network's performance by its own co-channel interference. In brief, Fig.~2 stipulates that the ASE of D$2$D networks  is jointly dependent on the
density of users and the propagation conditions.

Fig.~3 depicts the ASE of the
D$2$D underlay network as a function of the D$2$D mean access probability under Weibull fading. As observed in Fig.~3 the ASE is strongly coupled with the fading severity of the
propagation channel. For a D$2$D network more densely deployed than the cellular network
($\lambda_d > \lambda_c$), the fading severity $\nu_d$ plays a more important
role than  $\nu_c$. Hence, the attainable ASE is dramatically reduced when the fading severity
of inter-tier D$2$D  communication and cross-tier
interference channel is reduced. In fact,  a reduced power budget due to an increased fading parameter (small fading severity)
 outweighs the performance gain due to better propagation
conditions for the communication link.

Fig.~4 plots the ASE in Weibull fading for several
different values of $\eta$ against the D$2$D mean access probability. We observe that  reducing $\eta$ enlarges the D$2$D operational region in terms of access probability at smaller ASE values. This is in fact due to  less cross-iter interference with a reduced signal power at the D$2$D receiver. Consequently,
although a smaller $\eta$ may increase
the access probability limit,  the attained performance
may deteriorate due to the reduction of the overall ASE. 

\begin{figure}[h!]
\centering
\centerline{\includegraphics[ width=8cm]{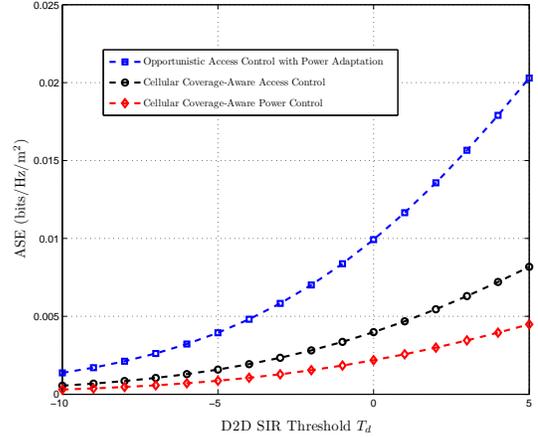}}
\caption{ ASE for D$2$D network according to different access and power
control methods under Weibull fading with $\lambda_d = 10^{-2}$, $\lambda_c = 10^{-3}$, $P_c = 1$, $\alpha=4$, $r_c =5$, $r_d=1$
$\rho^{c}_{th}= 0.7$,  $\nu_c = \nu_d = 2.5$, $T_c=10$ dB}\label{fig:BPSK}	
\end{figure}

Fig.~5, compares the performance of the cellular coverage-aware power or access control
and the opportunistic access control combined with cellular coverage-aware power control.
We notice that the latter scheme greatly improves the ASE
of D$2$D communications comparing to
sole adaptation of a single degree of freedom (transmit power or access probability) with an arbitrary
selection of the other resulting in a sub-optimal performance.
On the other hand,  the maximum ASE under cellular coverage-aware access control
 is higher than the one attained with power control.   However, the maximum ASE
is bounded by a wall due to the primary user's QoS
requirements.

\section{CONCLUSION}
 In this paper, we  developed a new methodology for modeling
and analyzing D$2$D-enabled cellular networks over general fading channels that relies on  the {\rm H}-transform theory.   This methodology subsumes
most known fading models and  more importantly enables
the unified analysis for the SINR  distribution of  D$2$D communications.  This framework is traditionally intractable due to the model-dependent limit on the distribution of
the SINR in previous derivations. We build upon the developed statistical machinery to formulate an  optimization scheme for D$2$D
networks in terms of SIR and spectral efficiency. This scheme  combines power control that is aware of
cellular coverage probability with opportunistic access control. That is  to
reduce the interference caused by D$2$D communications and
maximize the area spectral efficiency of D$2$D communications.  We show that  the optimal proportion and transmit power of active devices can be easily obtained by  simple fading model-specific formulas, thereby serving as a useful tool for network designers  to better understand and fine-tune  the performance of D$2$D-enabled cellular networks.

\section{Appendix}
\subsection{Proof of Theorem 1:}
The SINR CCDF  $\mathbb{P}^{x}(T)$  may be retrieved from its Laplace transform as
 \begin{equation}
{\cal L}_{\mathbb{P}^{x}}(z)=\frac{1}{z}-\frac{M_{\text{SINR}}^{x}(z)}{z}, \quad  z\in \mathbb{R}_{+},
\label{cdf1}
\end{equation}
where $M_{\text{SINR}}^{x}(z)$ denotes the SINR moment generating function recently derived  in \cite[Theorem 1]{trigui}. Hence it follows that
\begin{equation}
\mathbb{P}^{x}(T)=\!2\!\! \int_{0}^{\infty}\!\!\!\!\!{\cal E}_{h}\!\left[\sqrt{h}\Upsilon\right] {\cal E}_r\!\Bigg[\exp\left(-\frac{\sigma^{2}}{P}\xi^{2}r^{\alpha}\right){\cal L}_{{\cal I}_x}(\xi^{2}r^{\alpha})\Bigg] d\xi,
\label{p1}
\end{equation}
where ${\cal L}_{{\cal I}_x}(s)= {\cal E}\left[e^{-s I_x}\right]$ denotes the Laplace transform of the aggregate interference, and  $\Upsilon={\cal L}^{-1}\!\!\left(\!\frac{{\rm J}_1\left(2\sqrt{s h } \xi\right)}{\sqrt{s}}, T\!\right)$, where  ${\rm J}_{1}(\cdot)$ is the Bessel function of the second kind \cite[Eq. (8.402)]{grad} and ${\cal L}^{-1}(\cdot)$ stands for the inverse Laplace transform.
Resorting to \cite[Eq. (1.127) ]{H1} and \cite[Eq. (2.21)]{H1}, we get
\begin{equation}
\Upsilon=\frac{\xi \sqrt{h}}{T}{\rm H}_{1,2}^{1,0}\left[\frac{h \xi^{2}}{T}\!\Bigg|\begin{array}{ccc}(0,1) \\ (0,1),(-1,1)\end{array}\Bigg.\right].
\label{r1}
\end{equation}
Plugging (\ref{r1}) into (\ref{p1})  and carrying out the change of variable relabeling $\xi^{2}$  as $\xi$ yield
\begin{eqnarray}
\mathbb{P}^{x}(T)\!\!&=&\!\!\frac{1}{T} \int_{0}^{\infty}\!\!\!\!{\cal E}_{h}\!\!\left[h~{\rm H}_{1,2}^{1,0}\!\!\left[ \!\frac{h \xi}{T}\!\Bigg|\!\!\begin{array}{ccc}(0,1) \\ (0,1),(-1,1)\end{array}\Bigg. \!\!\right]\right]\nonumber \\ && {\cal E}_r\Bigg[\exp\left(-\frac{\sigma^{2}}{P}\xi^{2}r^{\alpha}\right){\cal L}_{{\cal I}}(\xi^{2}r^{\alpha})\Bigg] d\xi.
\label{pc1}
\end{eqnarray}

The Laplace transform of the interference at the cellular receiver, ${\cal L}_{{\cal I}_c}(s)$,
is evaluated as follows
\begin{equation}
{\cal L}_{{\cal I}_c}(s) = \exp(−2 \pi \lambda_c \Theta(s)),
\label{IL}
\end{equation}
where
\begin{eqnarray}
\Theta(s)&\overset{(a)}{=}&{\cal E}_{h}\left[\int_{r}^{\infty}\left(1-\exp\left(-s  h_c x^{-\alpha}\right)\right)x dx\right]\nonumber\\
&\overset{(b)}{=}&  s{\cal E}_{h}\left[h  \int_{r}^{\infty} x^{1-\alpha}e^{-s h x^{-\alpha}}{\rm {}_{\!1}F_{\!1}}\left(1,2,s h x^{-\alpha}\right )dx\right]\nonumber\\
&\overset{(c)}{=}& \frac{ s{\cal E}_{h} \left[h ~{\rm {}_{2}F_{2}}\left(1, \frac{2}{\alpha}+1; 2;\frac{2}{\alpha}+2; - s h r^{-\alpha}\right)\right] }{r^{\alpha-2}\alpha\left(1-\frac{2}{\alpha}\right)} ,
\label{agg}
\end{eqnarray}
and  the PGFL of a HPPP with intensity
function $\lambda_c$  is used in the first equality,  $(1-e^{-x})/x=  e^{-x}{\rm {}_{1}F_{1}}\left(1, 2; x\right)$ is applied in $(b)$,
and $(c)$ follows from letting $t=x^{-\alpha}$ and  applying $\int x^{\beta-1} e^{- c x}{\rm {}_{1}F_{1}}(a,b, c x)=\frac{x^{\beta}}{\beta}{\rm {}_{2}F_{2}}\left(b-a, \beta, b, \beta+1, -c x \right)$.

The Laplace transform of the interference at the probe D$2$D receiver, ${\cal L}_{{\cal I}_d}(\xi)$, can be evaluated as \cite[Eq. (22)]{d2dsku}
\begin{equation}
{\cal L}_{{\cal I}_d}(\xi)= \exp\left(- \pi  \lambda_d \xi^{\delta}\Gamma\left(1-\delta\right)  {\cal E}[h^{\delta}]\right).
\label{idp}
\end{equation}
Finally, plugging  (\ref{agg}) and (\ref{idp}) into (\ref{pc1}) with simple algebraic
manipulations, the final result presented in Theorem 1 follows.

\subsection{Proof of Proposition 1:}
 Let $h$ be a random variable with $f_{h}(y)=\frac{\left(\frac{m}{\Omega}\right)^{m}}{\Gamma(m)}x^{m-1} e^{- \frac{m}{\Omega} x}$ put in the form of a single H-variate \cite[Eq. (1.125)]{H1}, then
applying \cite[Eqs. (7.813), (9.31.5)]{grad} yields $\Psi(\xi,T)$ after some manipulations.
On the other hand, ${\cal A}^{d}_m(\xi, \alpha)$ is obtained from (\ref{Ax}) while ${\cal E}[h^{\delta}]=\frac{\Gamma(m+\delta)}{\Gamma(m)}\left(\frac{\Omega}{m }\right)^{\delta}$ and ${\cal A}^{c}_m(\xi, \alpha)$ follows from (\ref{Ax}) after applying \cite[Eq. (7.522.9) ]{grad}. Plugging all these results in  (\ref{GF}) and (\ref{Ax}) yields
\begin{eqnarray}
\mathbb{P}_{m}^{x}(T)&=&\frac{1}{\Gamma(m)}\int_{0}^{\infty}\xi^{-1}{\rm H}_{2,2}^{1,1}\left[\frac{ \Omega \xi}{ m T}\left|
\begin{array}{ccc}(1-m,1) \\(0,1) \end{array}\right. \right]\nonumber \\ &&\!\!\!\times {\cal E}_r\left[\exp\left(- \frac{\sigma^{2}}{P_x}\xi r^{\alpha}- {\cal A}_m^{x}\!\left(\xi r^{\alpha}, \delta\right)\right)\right]d\xi,
\label{COVm}
\end{eqnarray}
where
 \begin{eqnarray}
\left\{
  \begin{array}{ll}
   {\cal A}^{d}_m(\xi, \delta)=\pi \lambda_d \xi^{\delta}\left(\frac{\Omega}{ m}\right)^{\delta}\frac{\Gamma(1-\delta)\Gamma(m+\delta)}{\Gamma(m)},  \\
     {\cal A}^{c}_m(\xi, \delta)=\frac{\pi \delta \lambda_c \xi  \Omega~ {\rm {}_{3}F_{2}}\left( 1-\delta,1, m+1;2-\delta,2; - \frac{\Omega\xi }{m r^{2/\delta}} \right)}{ r^{2(1/\delta-1)}(1-\delta)}.
  \end{array}
\right.
\label{Am}
\end{eqnarray}
We assume that the distance  $r$  between a typical user and its associated MBS (cellular
link) or D$2$D helper (D$2$D link) follows a Rayleigh distribution, i.e. $f_{r_x}(r)=2 \pi \lambda_x r e^{-\pi \lambda_x r^{2}}, r\geq0, x\in\{c,d\}$ \cite{and}. Hence,  substituting $f_r(\cdot)$ for cellular and D$2$D users  in (\ref{COVm}) and resorting to \cite[Eq. (2.3)]{H1} yield (\ref{COVmf}) after some manipulations.

\subsection{Proof of Proposition 2:}
The proof follows from (\ref{GF}) with  $f_{h}(y)=\frac{\alpha}{\Phi}y^{\alpha-1} e^{- \frac{y^{\alpha}}{\Phi}}=\frac{\alpha}{\Phi}y^{\alpha-1}{\rm H}_{0,1}^{1,0}\!\left[ \frac{y^{\alpha}}{\Phi}\left|\!\!\!
\begin{array}{ccc} - \\ (0,1)\end{array}\right. \!\!\!\right]$
and applying \cite[Eqs. (2.3), (1.56)]{H1}. Besides,  ${\cal A}^{c}_{\cal W}(\xi, \alpha)$  follows by resorting to ${\cal E}[h^{\delta}]=\Gamma(1+\frac{\delta}{\alpha})\phi^{\frac{\delta}{\alpha}}$. On the other hand,  recalling that ${\rm {}_{p}F_{q}}(\underline{a}_p ; \underline{b}_{q}; -z)={\rm H}_{p,q+1}^{1,p}\!\bigg[ z\bigg|\!\!\!
\begin{array}{ccc} (1- \underline{a}_{p},1)\\ (1,1),  (1- \underline{b}_{q},1)\end{array}\bigg. \!\!\!\bigg]$  and applying \cite[Eqs. (2.3)]{H1}  yield ${\cal A}^{c}_{\cal W}(\xi, \alpha)$ after some manipulations using the Fox-H function properties in  \cite[Eqs. (1.2.3), (1.2.4)]{matai}. Plugging all these results in  (\ref{GF}) and (\ref{Ax}) yields
\begin{eqnarray}
\mathbb{P}_{\cal W}^{x}(T)\!\!\!&\!\!\!\!\!=\!\!\!\!\!&\!\!\!\frac{\nu}{ T \Phi} \int_{0}^{\infty}\left(\frac{T}{\xi}\right)^{\nu+1}{\rm H}_{1,1}^{1,0}\!\left[ \left(\frac{ T}{ \xi }\right)^{\nu}\frac{1}{\Phi}\left|\!\!\!
\begin{array}{ccc} (1-\nu,\nu) \\ (0,1)\end{array}\right. \!\!\!\right]\nonumber \\ && \times {\cal E}_r\left[\exp\left(- \frac{\sigma^{2}}{P_x} \xi r^{\alpha} -{\cal A}_{\cal W}^{x}(\xi r^{\alpha}, \delta)\right)\right]d\xi,
\label{COVw}
\end{eqnarray}
where
\begin{equation}
\!\!\!\left\{
  \begin{array}{ll}
   \!\!\!\!{\cal A}_{\cal W}^{d}(\xi, \delta)\!=\! \pi \lambda_d \xi^{\delta}\Phi^{\frac{\delta}{\nu}}\Gamma\left(1-\delta\right)\Gamma\left(1+\frac{\delta}{\nu}\right), \\
     \!\!\!\!{\cal A}_{\cal W}^{c}(\xi, \delta)\!=\!\frac{\delta \lambda_c r^{2}}{1-\delta} {\rm H}_{3,3}^{3,1}\left[\frac{  r^{\frac{2}{\delta}}}{\xi \Phi^{\frac{1}{\nu}} }\left|
\begin{array}{ccc}\!\!(-1,1), (1-\delta,1),(1,1)\\ (1,\frac{1}{\nu}),(-\delta,1), (0,1)\end{array}\right. \!\!\!\right].
  \end{array}
\right.
\label{Axw}
\end{equation}
Finally, substituting $f_r(\cdot)$ for cellular and D$2$D users  and proceeding
as before completes the proof.

\subsection{$\kappa$-$\mu$ and Shadowed $\kappa$-$\mu$ As
Special Cases of the Degree-2 Fox's H-Function Fading Model }

The $\kappa$-$\mu$
distribution, first introduced in \cite{kmu},  can be regarded as a generalization of the classic
Rician fading model for LOS scenarios.
Let $h$ be a random variable statistically following a  $\kappa$-$\mu$
 distribution \cite{unif} with mean $\Omega={\cal E}[h]$ and non-negative
real shape parameters $\kappa$, and  $\mu$, with
\begin{equation}
f_{h}(x)=\frac{ \mu \left(\frac{1+\kappa}{\Omega}\right)^{\frac{\mu+1}{2}}x^{\frac{\mu-1}{2}}e^{-\frac{ \mu(1+\kappa)x}{\Omega}  }}{e^{\kappa\mu} \kappa^{\frac{\mu-1}{2}}}{\rm I}_{\mu-1}
\left(2\mu\sqrt{\frac{ \kappa(1\!+\!\kappa)}{\Omega} x }\right),
\label{fiku}
\end{equation}
where ${\rm I}_{b}(\cdot)$  stands for the modified Bessel function of the first kind of order $b$ \cite[Eq. (8.431.1)]{grad}.
Recognizing that  \cite[A.7 ]{H1} \begin{equation}
{\rm I}_{\nu}\left(z\right)= i^{-\nu} {\rm H}_{0,2}^{1,0}\bigg[ -\frac{z^{2}}{4}\bigg|\left(\frac{\nu}{2},1\right), \left(-\frac{\nu}{2},1\right)\bigg. \bigg],
\label{Iku}
\end{equation}
 where $i^{2}=-1$, then from (\ref{GF})  it follows that in $\kappa$-$\mu$ fading
\begin{eqnarray}
\!\Psi(\xi,T)\!\!\!&=&\!\!\! {{\cal C}}_{\kappa\mu}\int_{0}^{\infty}x^{\frac{\mu+1}{2}}e^{-\frac{ \mu(1+\kappa)x}{\Omega}  }\nonumber\\ &&\!\!\!\!\!\!\!\!\!\!\!\!\!\!\!\!\!\!\! \times {\rm H}_{1,2}^{1,0}\left[\frac{    \xi  x}{T }\left|
\begin{array}{ccc}(0,1) \\ (0,1),(-1,1) \end{array}\right. \right]\nonumber\\ && \!\!\!\!\!\!\!\!\!\!\!\!\!\!\!\!\!\! \times {\rm H}_{0,2}^{1,0}\bigg[ -\frac{ \mu^{2}\kappa(1\!+\!\kappa)}{\Omega} x\bigg|\left(\frac{\nu}{2},1\right), \left(-\frac{\nu}{2},1\right)\bigg. \bigg]dx.
\end{eqnarray}
The last H-transform is known as the Laplace transform of two  Fox's-H function given by \cite[Eq. (2.6.2)]{matai} as
\begin{eqnarray}
\!\Psi(\xi,T)\!&=&\!{{\cal C}}_{\kappa\mu}\left(\frac{ \mu(1\!+\!\kappa)}{\Omega}\right)^{-\frac{\mu+3}{2}}\nonumber \\ && \!\!\!\!\!\!\!\!\!\!\!\!\!\!\!\!\!\!\!\!\!\!\!\!\!\!\!\!\!\!\!\!\!\!\!\!\!\!\!\!\!\!
{\rm H}_{1,[0,1],0,[2,2]}^{1,0,0,1,1}\!\! \left[\!\!{- \mu \kappa\atop \frac{    \xi \Omega}{T \mu (1\!+\!\kappa)}} \!\!\! \left| \begin{array}{cccc}\left(1\!+\!\frac{\mu+1}{2},1\right) \!\\-;\!(0,1)\\ \!\!\! \left(\frac{\mu\!-\!1}{2},1\right)\!,\!\left(\frac{1\!-\!\mu}{2},1\right)\!;\!(0,1)\!,\!(-1,1)\end{array}\right.\!\!\!\! \right],
\end{eqnarray}
where ${\rm H}[\cdot, \cdot]$
denotes the generalized Fox's H-function of two variables \cite[Eq. (1.1)]{mittal} and it
reduces with the help of  \cite[Eq. (2.3.1)]{matai} to the generalized Meijer's G-function of two variables. Recalling that  ${\cal E}[h^{j}]=\frac{\left(\frac{\Omega}{\mu(1+\kappa)}\right)^{j}\Gamma(\mu+j)}{e^{\mu \kappa}}{\rm {}_{1\!}F_{\!1}}\left(\mu+j, \mu;-\mu \kappa\right)$ under  $\kappa$-$\mu$ fading  \cite[Eq. (10)]{d2dsku}, thereby yielding ${\cal A}_{ \kappa\mu}^{d}(\xi, \alpha)$  as in (\ref{Axkmu}). On the other hand ${\cal A}_{ \kappa\mu}^{c}(\xi, \alpha)$  is obtained  from (\ref{Ax}) using ${\rm {}_{2}F_{2}}\left( 1-\delta,1;2-\delta,2; - \xi h \right)=\sum_{k=1}^{2}{\rm {}_{1\!}F_{\!1}}\left(a_k, a_k+1; - \xi h \right)\prod_{j=1, j\neq k}^{3 }\frac{a_j}{a_j-a_k}$ where $a_k \in \{1, 1-\delta\}, k=1,\ldots,2$ then applying \cite[Eq. (27)]{humbert} yield
\begin{eqnarray}
{\rm \Psi}_{1}\!\!\left(\!a,b;c,c';w,z\!\right)\!\!\!\!\!&=&\!\!\!\!\!\frac{\Gamma(c')}{\Gamma(a)}z^{\frac{1-c'}{2}}\int_{0}^{\infty}\!\!\!\!\!t^{a-\frac{1+c'}{2}}e^{-t}{\rm I}_{c'-1}(2\sqrt{ t z})\nonumber \\ && {\rm {}_{1}F_{1}}\left(b, c ,   w t \right)dt,
\end{eqnarray}
where ${\rm \Psi_1}(\cdot,\cdot;\cdot,\cdot;\cdot,\cdot)$   stands for the Humbert function of the first kind \cite[Eq. (2)]{humbert}.
Plugging all these resulst into (\ref{GF}) yields the D$2$D and cellular  CCDFs  as
\begin{eqnarray}
\mathbb{P}_{\kappa\mu}^{x}(T)\!\!&\!\!\!\!\!=\!\!\!\!\!&\!\!\frac{{\cal\widetilde{ C}}_{\kappa\mu}}{T}\int_{0}^{\infty}\!\!\!\!{\rm G}_{1,[0,1],0,[2,2]}^{1,0,0,1,1}\!\! \left[\!\!{- \mu \kappa\atop \frac{    \xi \Omega}{T \mu (1\!+\!\kappa)}}  \left| \begin{array}{cccc}1\!+\!\frac{\mu+1}{2} \\-;\!0\\ \frac{\mu-1}{2},\frac{1-\mu}{2};0,\!-1\end{array}\right.\!\! \right]\nonumber\\ &&\times {\cal E}_r\left[\exp\left(\!-\frac{\sigma^{2}}{P_x} \xi r^{\alpha} \!-\!\! {\cal A}_{\kappa\mu}^{x}(\xi r^{\alpha}, \delta)\!\right)\right]d\xi,
\label{COVku}
\end{eqnarray}
where ${\cal \widetilde{C}}_{\kappa\mu}=\frac{ \mu (1+\kappa)^{\frac{\mu+1}{2}}}{e^{\kappa\mu}\Omega \kappa^{\frac{\mu-1}{2}}}$, and  ${\cal A}_{\kappa\mu}^{x}(\xi, \alpha)$ is obtained as
\begin{equation}
\!\!\left\{
  \begin{array}{ll}
  \!\!\!{\cal A}_{\kappa\mu}^{d}(\xi, \delta)\!=\!\frac{\pi \lambda_d \xi^{\delta}\Gamma(1-\delta)\left(\frac{\Omega}{\mu(1+\kappa)}\right)^{\delta}\Gamma(\mu+\delta){\rm {}_{1\!}F_{\!1}}\left(\mu+\delta, \mu;-\mu \kappa\right)}{\Gamma(\mu)}, \\
    \!\!\!{\cal A}_{\kappa\mu}^{c}(\xi, \delta)\!=\!\frac{\delta \lambda_c \Omega \xi e^{-\mu \kappa}\sum_{k=1}^{2}\Theta_k {\rm \Psi}_1\left(\mu+1, a_k; a_k+1, \mu; \mu \kappa, -\frac{\xi \Omega}{\mu (1+\kappa)}\right) }{r^{2(1/\delta-1)}(1-\delta)},
  \end{array}
\right.
\label{Axkmu}
\end{equation}
where  $\Theta_k=\prod_{j=1, j\neq k}^{2}\frac{a_j}{a_j-a_k} $ with $a_k \in \{1, 1-\delta\}, k=1,\ldots,2$, and ${\rm G}_{a,[c,e],b,[d,f]}^{p,q,k,r,l}[\cdot,\cdot]$  is   the generalized Meijer's G-function  of two variables  \cite{verma}.

 In interference-limited $\kappa$-$\mu$  environment, the SIR CCDF of D$2$D links is obtained as
\begin{eqnarray}
\mathbb{P}_{\kappa\mu}^{d}(T)
\!\!\!&\!\!\!\!\!\overset{(a)}{=}\!\!\!\!\!&\!\!\!\frac{{\cal\widetilde{ C}}_{\kappa\mu}}{T}\int_{0}^{\infty}\!\!\!\!{\rm G}_{1,[0,1],0,[2,2]}^{1,0,0,1,1}\!\! \left[\!\!{- \mu \kappa\atop \frac{    \xi \Omega}{T \mu (1\!+\!\kappa)}}  \left| \begin{array}{cccc}1\!+\!\frac{\mu+1}{2} \\-;\!0\\ \frac{\mu-1}{2},\frac{1-\mu}{2};0,\!-1\end{array}\right.\!\! \right]\nonumber\\ &&\times {\cal E}_r\left[{\rm H}_{0,1}^{1,0}\left[{\cal A}_{\kappa\mu}^{d}(\xi r^{\alpha}, \delta)\left|\!\!\!
\begin{array}{ccc} - \\ (0,1)\end{array}\right. \!\!\!\right]\right]d\xi\nonumber \\ &&
=\delta \widetilde{{\cal C}}_{\kappa\mu}\frac{\left(\lambda_d \kappa_{\kappa\mu}\right)^{-\frac{1}{\delta}} }{ T} {\cal E}_r \Bigg[ r^{2}\nonumber \\ && \!\!\!\!\!\!\!\!\!\!\!\!\!\!\!\!\!\!\!\!\!\!\!\!\!\!\!\!\!\!\!\!H_{0,[2,2],1,[0,1]}^{1,1,2,0,0}\!\! \left[\!\!{(-
\mu \kappa)^{-\delta}\atop \lambda_d \kappa_{\kappa\mu} T^{\delta} r^{2}} \Bigg| \!\!\!\begin{array}{cccc}(-\frac{\mu\!+\!1}{2},\delta) \\-;(\frac{1}{\delta},1);(1,\delta)\\ (\frac{3\!-\!\mu}{2},\delta), (\frac{1\!+\!\mu}{2},\delta);(1,\delta),(2,\delta)\end{array}\Bigg.\!\!\! \right]\!\!\Bigg],
\label{pkuh}
\end{eqnarray}
where $(a)$ follows from substituting ${\cal A}_{\kappa\mu}^{d}(\xi r^{\alpha}, \delta)$ by its expression in (\ref{Axkmu}) after recognizing the Fox's-H representation of the exponential function \cite[Eq. (1.7.2)]{matai}   and employing  \cite[Eq. (2.11)]{matai}. Moreover,  $\kappa_{\kappa\mu}=\frac{\Gamma(1\!-\!\delta) \Gamma(\mu+\delta)e^{-\mu\kappa }{\rm {}_{1\!}F_{\!1}}\left(\mu\!+\!\delta, \mu;-\mu \kappa\right)}{\Gamma(\mu)}$. Notice that the $\kappa$-$\mu$ includes the Rayleigh $(\kappa \rightarrow 0, \mu = 1)$,  Nakagami-$m$ $(\kappa \rightarrow 0, \mu =m)$, and Rician  $(\kappa=K, \mu = 1)$ fading models as special cases, where $K$ is the Rician factor.

In shadowed $\kappa$-$\mu$   distribution the dominant signal components  are  subject to Nakagami-$m$ shadowing with pdf \cite[Table I]{paris}
\begin{eqnarray}
f_{h, S\kappa-\mu}(y)\!\!\!&=&\!\!\!\frac{ \mu^{\mu} m^{m}(1+\kappa)^{\mu}}{\Gamma(\mu)\Omega^{\mu} (\mu \kappa+m)^{m}}\left(\frac{y}{\Omega}\right)^{\mu-1}e^{-\frac{ \mu(1+\kappa)}{\Omega} y }\nonumber\\ &&{\rm {}_{1\!}F_{\!1}}\left(m, \nu , \frac{ \mu^{2} \kappa(1+\kappa)}{\Omega(\mu \kappa+m)} y \right),
\label{pdfsku}
\end{eqnarray}
where  ${\rm {}_{1\!}F_{\!1}}(\cdot)$ denotes the confluent hypergeometric function of \cite[Eq. (13.1.2)]{grad}.  Recalling that
\begin{equation}
{\rm {}_{1\!}F_{\!1}}\left(a,b;z\right)=\frac{\Gamma(b)}{\Gamma(a)} {\rm H}_{1,2}^{1,1}\bigg[ -z\bigg|\begin{array}{ccc}(1-a,1) \\ (0,1),(1-b,1)\end{array}\bigg. \bigg],
\end{equation}
then the D$2$D and cellular SINR CCDFs follow along the same line of (\ref{COVku}) as
\begin{eqnarray}
\mathbb{P}_{S\kappa\mu}^{x}(T)\!\!\!&\!\!\!\!\!=\!\!\!\!\!&\!\!\!\frac{\widetilde{{\cal C}}_{S\kappa\mu}}{T}\!\!\int_{0}^{\infty}\!\!\!\!\!\!G_{1,[1,1],0,[2,2]}^{1,1,0,1,1} \left[\!\!\!{-\frac{ \mu \kappa}{(\mu \kappa\!+\!m)}\atop \frac{    \xi \Omega}{T \mu (1\!+\!\kappa)}}  \left| \begin{array}{cccc}1+\mu \\1-m;0\\ 0,1-\mu;0,-1\end{array}\right.\!\!\! \right]\nonumber\\ &&\times {\cal E}_r\left[\exp\left(\!-\frac{\sigma^{2}}{P_x} \xi r^{\alpha} \!-\!\! {\cal A}_{S\kappa\mu}^{x}(\xi r^{\alpha}, \delta)\!\right)\right]d\xi,
\label{COVsku}
\end{eqnarray}
where ${\cal \widetilde{C}}_{S\kappa\mu}=\frac{ \frac{\Omega}{\mu}}{\Gamma(m)(1+\kappa)\left(\frac{\mu \kappa}{m}+1\right)^{m}}$. Moreover in (\ref{COVsku}), ${\cal A}_{S\kappa\mu}^{x}(\xi, \alpha)$ is obtained as
\begin{equation}
\left\{
  \begin{array}{ll}
  \!\!\!\!\!{\cal A}_{S\kappa\mu}^{d}(\xi, \delta)\!\overset{(a)}{=}\!\frac{\pi \lambda_d \xi^{\delta}\Gamma(1\!-\!\delta)\Gamma(\mu+\delta){\rm {}_{2\!}F_{\!1}}\left(\mu\!-\!m,\mu\!+\!\delta, \mu;-\frac{\mu \kappa}{m}\right)}{\left(\frac{\Omega}{\mu(1\!+\!\kappa)}\right)^{-\delta}\left(\frac{\mu \kappa}{m}+1\right)^{m-\mu-\delta}\Gamma(\mu)}; \\
     \!\!\!\!\!{\cal A}_{S\kappa\mu}^{c}\!(\xi,\delta)\!\overset{(b)}{=}\!\frac{\delta \lambda_c \Omega\xi  \sum_{k=1}^{2}\!\Theta_k{\rm F}_2\!\left(\mu\!+\!1, a_k, m;a_k\!+\!1, \mu; \frac{\mu \kappa}{\mu \kappa+m},\frac{- \Omega\xi r^{\!\frac{-2}{\delta}}}{\mu (1+\kappa)}\right) }{r^{2(1/\delta-1)}(1+\kappa)(1-\delta)(\frac{\mu \kappa}{m}+1)^{m}},
  \end{array}
\right.
\label{Axskmu}
\end{equation}
where $(a)$ follows after recognizing that ${\cal E}[h^{j}]=\frac{\left(\frac{\Omega}{\mu(1+\kappa)}\right)^{j}\Gamma(\mu+j)}{\left(\frac{\mu \kappa}{m}+1\right)^{m-\mu-j}\Gamma(\mu)}{\rm {}_{2\!}F_{\!1}}\left(\mu-m,\mu+j, \mu;-\frac{\mu \kappa}{m}\right)$ \cite[Eq. (10)]{d2dsku} thereby yielding ${\cal A}_{ S\kappa\mu}^{d}(\xi, \alpha)$. On the other hand, ${\cal A}_{ S\kappa\mu}^{c}(\xi, \alpha)$  is obtained  from (\ref{Ax}) along the same line of (\ref{Axkmu}) while considering the following integral form
\begin{eqnarray}
{\rm F}_{2}\!\!\left(\!a,b,b',c,c';\frac{w}{p},\frac{ z}{p}\!\right)\!\!\!\!\!&=&\!\!\!\!\!\frac{p^{a}}{\Gamma(a)}\int_{0}^{\infty}\!\!\!\!\!x^{a-1}e^{- p x}{\rm {}_{1}F_{1}}\left(b, c ,   w x \right)\nonumber \\ &&{\rm {}_{1}F_{1}}\left(b', c' ,    x z \right)dx,
\end{eqnarray}
where ${\rm F_{2}}(a,b, b';c , c'; x , y)$ stands for the Appell's hypergeometric function of the second kind  \cite[Eq. (27)]{appel}.

 We assume that the communication
is interference limited and hence thermal noise is
negligible. Then the coverage of D$2$D communication in $\kappa$-$\mu$ shadowed fading is obtained as
\begin{eqnarray}
\mathbb{P}_{S\kappa\mu}^{d}(T)\!\!\!&\!\!\!\!\!=\!\!\!\!\!&\!\!\!\frac{\widetilde{{\cal C}}_{S\kappa\mu}}{T}\!\!\! \int_{0}^{\infty}\!\!\!\!\!\!G_{1,[1,1],0,[2,2]}^{1,1,0,1,1} \left[\!\!\!{-\frac{ \mu \kappa}{(\mu \kappa\!+\!m)}\atop \frac{    \xi \Omega}{T \mu (1\!+\!\kappa)}}  \left| \begin{array}{cccc}1+\mu \\1-m;0\\ 0,1-\mu;0,-1\end{array}\right.\!\!\! \right]\nonumber \\ && \times {\cal E}_r\left[{\rm H}_{0,1}^{1,0}\left[{\cal A}_{S\kappa\mu}^{d}(\xi r^{\alpha}, \delta)\left|\!\!\!
\begin{array}{ccc} - \\ (0,1)\end{array}\right. \!\!\!\right]\right]d\xi\nonumber \\\!\!\!&\!\!\!\!\!=\!\!\!\!\!&\!\!\!\delta \widetilde{{\cal C}}_{S\kappa\mu}\frac{\left(\lambda_d \kappa_{S\kappa\mu}\right)^{-\frac{1}{\delta}} }{ T} {\cal E}_r \Bigg[ r^{2}\nonumber \\ && \!\!\!\!\!\!\!\!\!\!\!\!\!\!\!\!\!\!\!\!\!\!\!\!\!\!\!\!\!\!H_{0,[2,2],1,[1,1]}^{1,1,2,1,0}\!\! \Bigg[\!\!{\left(-\frac{\mu\kappa+ m}{\mu \kappa}\right)^{\delta}\atop \lambda_d \kappa_{S\kappa\mu} T^{\delta} r^{2} } \!\! \left| \!\!\!\begin{array}{cccc}(-\mu,\delta) \\(\!m,\delta);(\frac{1}{\delta},1);(1,\delta)\\ (1,\delta), (\mu,\delta);(1,\delta),(0,\delta)\end{array}\right.\!\!\! \Bigg]\Bigg],
\label{SKUlim}
\end{eqnarray}
where $\kappa_{S\kappa\mu}=\frac{\Gamma(1\!-\!\delta) \Gamma(\mu+\delta){\rm {}_{2\!}F_{\!1}}\left(\mu\!-\!m,\mu\!+\!\delta, \mu;-\frac{\mu \kappa}{m}\right)}{\left(\frac{\mu \kappa}{m}+1\right)^{m-\mu-\delta}\Gamma(\mu)}$.

\end{document}